\documentclass[10pt,twocolumn,superscriptaddress,nofootinbib,aps,floatfix,amsfonts,preprintnumbers,amsmath,amssymb,groupedaddress,prd]{revtex4-1}

\usepackage{graphicx}
\usepackage[colorlinks,pdfusetitle]{hyperref}

\newcommand{\orcid}[1]{\href{https://orcid.org/#1}{#1}}
\newcommand{\wh}[1]{\widehat{#1}}
\newcommand{\Dmsqee}{\Delta m^2_{ee}}
\newcommand{\eps}{\epsilon}
\newcommand{\e}[1]{\times10^{#1}}
\newcommand{\Ssol}{\mathcal S_{\odot}}
\newcommand{\Satm}{\mathcal S_{\rm atm}}

  \newcommand{\dmsqee}{\Delta m^2_{ee}} 
  
  \newcommand{\dmsqatm}{\Delta m^2_{31}}
   \newcommand{\dmsqatn}{\Delta m^2_{32}}

  \newcommand{\dmsqss}{\Delta m^2_{21}}
  
  \newcommand{\ctot}{\cos 2 \theta_{12}}

\begin{document}

\title{Simple and Precise Factorization of the Jarlskog Invariant\\for Neutrino Oscillations in Matter}

\author{Peter B.~Denton}
\email{pdenton@bnl.gov}
\thanks{\orcid{0000-0002-5209-872X}}
\affiliation{Physics Department, Brookhaven National Laboratory, Upton, New York 11973, USA}

\author{Stephen J.~Parke}
\email{parke@fnal.gov}
\thanks{\orcid{0000-0003-2028-6782}}
\affiliation{Theoretical Physics Department, Fermi National Accelerator Laboratory, Batavia, IL 60510, USA}

\begin{abstract}
For neutrino propagation in matter, we show that the Jarlskog invariant, which controls the size of true CP violation in neutrino oscillation appearance experiments, factorizes into three pieces: the vacuum Jarlskog invariant times two simple two-flavor matter resonance factors that control the matter effects for the solar and atmospheric resonances independently.
If the solar effective matter potential and the atmospheric effective $\Delta m^2$ are chosen carefully for these two resonance factors, then the fractional corrections to this factorization are an impressive 0.04\% or smaller.
We also show that the inverse of the square of the Jarlskog in matter ($1/\widehat{J}^2$) is a fourth order polynomial in the matter potential which guarantees that it can be factored into two quadratics which immediately implies the functional form of our approximate, factorized expression.
\end{abstract}

\preprint{FERMILAB-PUB-19-072-T}


\maketitle

\section{Introduction}
\label{sec:introduction}
The discovery of an invariant, the Jarlskog invariant \cite{Jarlskog:1985ht}, that controls the size of CP violation in both quark and lepton sectors was a monumental step in the understanding of flavor physics.
For neutrinos, using the standard parameterization of the PMNS matrix \cite{Maki:1962mu,Pontecorvo:1967fh}, the Jarlskog invariant
 is given by
\begin{equation}
J\equiv s_{13}c^2_{13}s_{12}c_{12}s_{23}c_{23}\sin\delta\,,
\label{eq:Jvac}
\end{equation}
where we use the usual notation, $c_{ij}=\cos\theta_{ij}$, $s_{ij}=\sin\theta_{ij}$, and $\delta$ is the CP-violating phase.
The CP-violating part of the vacuum neutrino oscillation probability in the appearance channels, e.g.~$\nu_\mu \rightarrow \nu_e$, is given by \cite{Bilenky:1987ty}
\begin{equation}
8 J\sin \Delta_{31} \sin \Delta_{32} \sin \Delta_{21}\,,
\label{eq:CPvac}
\end{equation}
where the kinematic phases are given by $\Delta_{jk} = \Delta m^2_{jk} L/4E_\nu$ with $\Delta m^2_{jk} = m^2_j -m^2_k$ for an experiment of baseline $L$ and neutrino energy $E_\nu$.

For neutrinos propagating in matter,
as in the NOvA \cite{Ayres:2004js}, T2K \cite{Itow:2001ee}, DUNE \cite{Acciarri:2016crz} and T2HK(K) \cite{Abe:2014oxa,Abe:2016ero} experiments,
the part of the appearance  oscillation probability that depends on the intrinsic CP violation is given by
\begin{equation}
8 \wh{J}\sin \wh{\Delta}_{31} \sin \wh{\Delta}_{32} \sin \wh{\Delta}_{21}\,,
\label{eq:CPmat}
\end{equation}
where $\wh{x}$ is the matter value for the vacuum variable $x$.
The Jarlskog invariant in matter, $\wh{J}$, is given by same expression as eq.~\ref{eq:Jvac}, but with the mixing angles and phase replaced by their matter values \cite{Zaglauer:1988gz,Krastev:1988yu,Parke:2000hu}.
Both $\theta_{12}$ and $\theta_{13}$ have a strong dependence on density of the matter and the energy of the neutrino through the Wolfenstein matter potential \cite{Wolfenstein:1977ue}, $a$, given by
\begin{equation}
a \equiv 2\sqrt{2} G_F N_e E_\nu\,,
\end{equation}
where
$G_F$ is the Fermi constant, $N_e$ is the number density of electrons, and $E_\nu$ is the neutrino energy in the matter rest frame.

\section{The Approximate Factorization}
\label{sec:result}
While the exact expressions for the mixing angles in matter are extremely complicated \cite{Zaglauer:1988gz}, it is possible to relate the Jarlskog invariant in matter to the vacuum Jarlskog, at the 0.04\% level, as simply
\begin{equation}
J\approx\Ssol\, \Satm\,\wh{J} \,,
\label{eq:Jmat}
\end{equation}
where
\begin{align}
\Ssol&=\sqrt{(\cos2\theta_{12}-c_{13}^2a/\Delta m^2_{21})^2+\sin^22\theta_{12}}\,, \nonumber \\
\Satm&=\sqrt{(\cos2\theta_{13}-a/\Dmsqee)^2+\sin^22\theta_{13}}\,.
\label{eq:S13,S12}
\end{align}
Eq.~\ref{eq:Jmat} shows simply how to relate the quantity measured in experiments, $\wh J$, to the amount of CP violation in the lepton sector, $J$.
The $\mathcal S$ factors are the two-flavor resonance factors associated with the solar and atmospheric resonances.
Like the Jarlskog invariant, these $\mathcal S$ factors can also be written in a convention independent form, see eq.~\ref{eq:indep}.

The precision scales like
 ${\mathcal O}(s_{13}^2\cos2\theta_{12} (\Delta m^2_{21}/\Delta m^2_{ee}))$ and ${\mathcal O}(s^2_{12}c^2_{12} (\Delta m^2_{21}/\Delta m^2_{ee})^2)$
  leading to an actual fractional precision of $\sim0.04\%$ for this factorization.
To achieve this level of precision, we note that the following  are crucial:
\begin{itemize}
\item for the solar (1-2) resonance factor, $\Ssol$, the effective matter potential is $c_{13}^2a$, not just $a$,
\item for the atmospheric (1-3) resonance factor, $\Satm$, the effective $\Delta m^2$ is\\[1mm]
$\Delta m^2_{ee} \equiv c_{12}^2 \Delta m^2_{31}+s_{12}^2\Delta m^2_{32}$ \cite{Nunokawa:2005nx,Parke:2016joa},
not $\Delta m^2_{31(2)}$.
\end{itemize}
\begin{figure}
\centering
\includegraphics[width=\columnwidth]{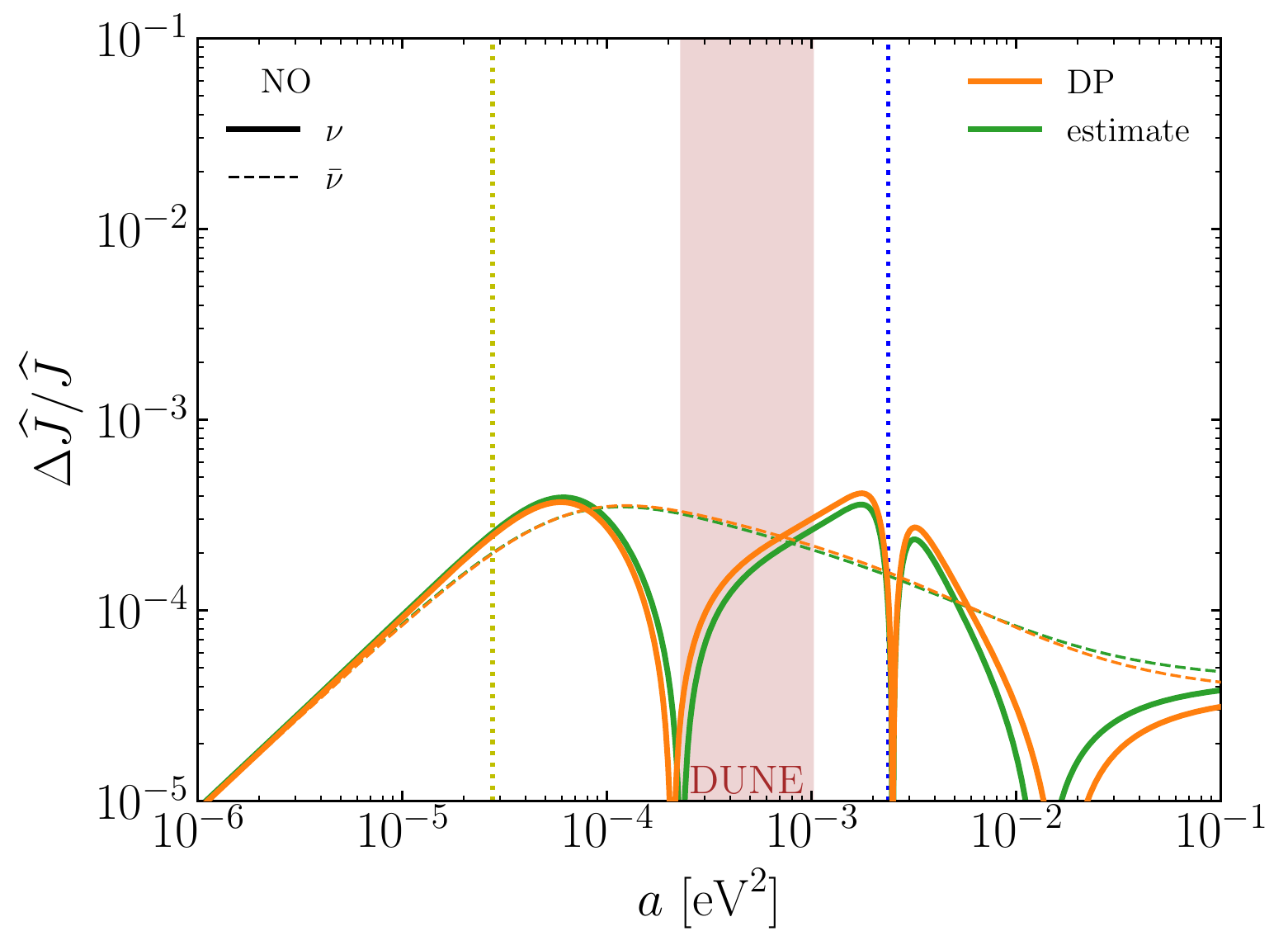}
\caption{The fractional precision in $\wh J$ compares our approximate expression with the exact expression calculated from \cite{Zaglauer:1988gz}, or eqs.~\ref{eq:evaluefn} and \ref{eq:Dprod}.
The orange curves our approximate expression from eqs.~\ref{eq:Jmat} and \ref{eq:S13,S12}.
The green curves are the analytic approximation of the precision shown in eq.~\ref{eq:Jh error}.
The yellow and blue vertical lines are the solar and atmospheric resonances respectively.
The vertical strip is the amount of matter potential that DUNE will probe.
The downward spikes occur where the exact and approximate expressions cross.
The normal mass ordering (NO) is assumed.}
\label{fig:error estimate}
\end{figure}
In fig.~\ref{fig:error estimate}, we have plotted the fractional precision to the approximation in eq.~\ref{eq:Jmat} as a function of the matter potential for both neutrinos and anti-neutrinos and find that the expression is precise to the 0.04\% level or better.
We have numerically verified that $c_{13}^2$ is the optimal correction, $\Dmsqee$ is the optimal atmospheric mass splitting, and that these results are generally independent of the mass ordering.

\begin{figure}
\centering
\includegraphics[width=\columnwidth]{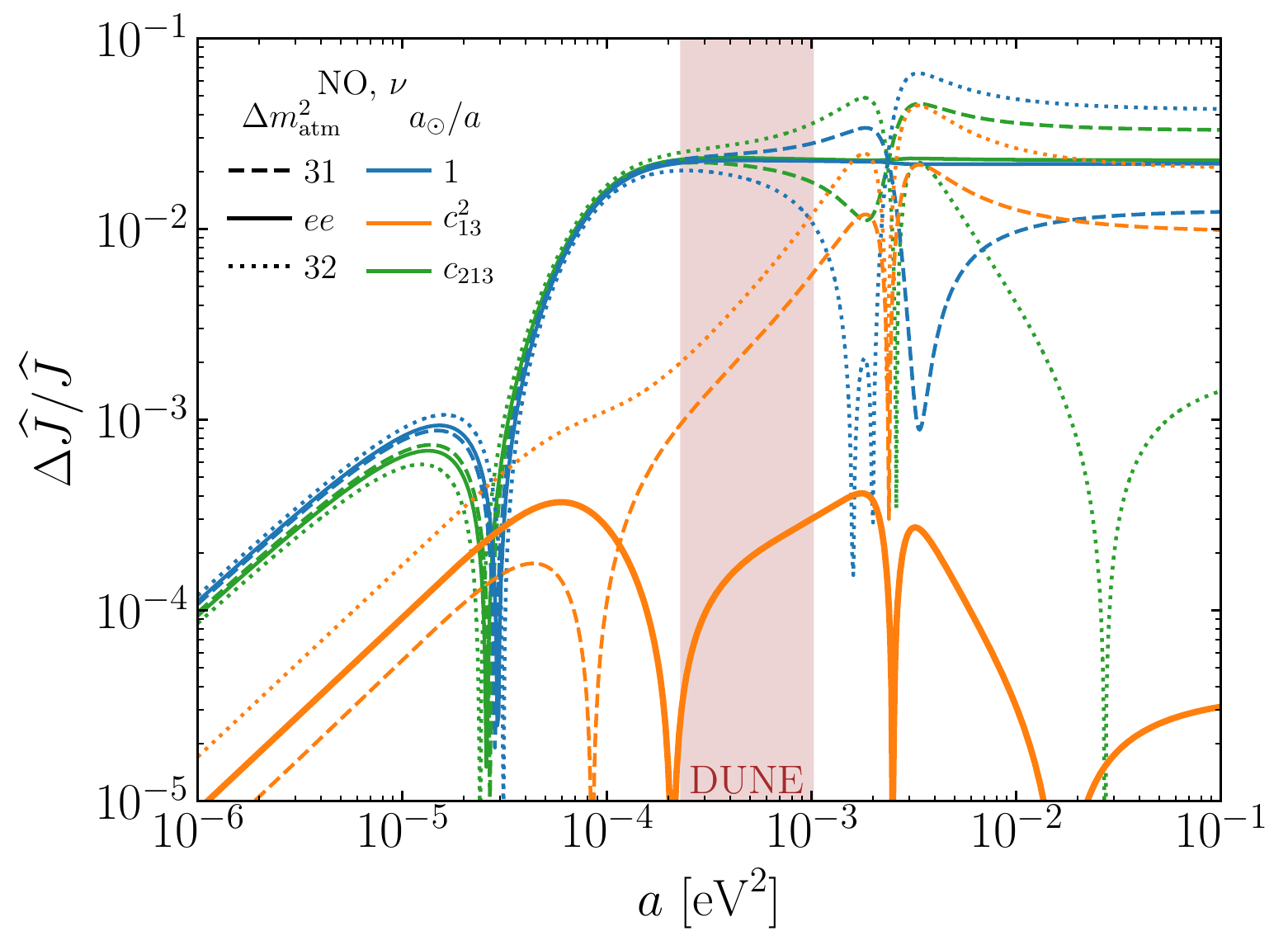}
\caption{The figure is the same as fig.~\ref{fig:error estimate} for neutrinos only.
We now vary the $\Delta m^2$ that appears in $\mathcal S_{\rm atm}$ and the correction to the matter potential that appears in $\mathcal S_\odot$.
Our solution is shown in the solid orange curve and is clearly the most precise in general and, in particular, for DUNE's region of interest.
The result from \cite{Wang:2019yfp} is the blue solid curve.}
\label{fig:error comparison}
\end{figure}

Without the $c_{13}^2$ term in $\Ssol$, or for different values of the solar matter potential, the precision is 2-3\% independent of the atmospheric mass splitting used in $\Satm$.
The precision is $\mathcal O(s_{13}^2)$ and $\mathcal O(\frac{\Delta m^2_{21}}{\Dmsqee})$, see the expression in ref.~\cite{Wang:2019yfp}.
With the $c_{13}^2$ term the precision is better, but still at the 2\% level, although it is better in DUNE's region of interest, down to $\sim0.1\%-1\%$, depending on which atmospheric $\Delta m^2$ is used 
 in the $\Satm$ term.
When the atmospheric splitting is $\Dmsqee$, the precision improves considerably down to the $0.04\%$ level or better for \emph{any matter potential}.
That is, the precision is ${\mathcal O}(s_{13}^2\cos2\theta_{12}(\Delta m^2_{21}/\Delta m^2_{ee}))$, ${\mathcal O}(s^2_{12}c^2_{12}(\Delta m^2_{21}/\Delta m^2_{ee})^2)$, or better for all values of the matter potential.
This is all shown in fig.~\ref{fig:error comparison}.
Our result is the solid orange curve, $(ee,c^2_{13})$ and the result from ref.~\cite{Wang:2019yfp} which is the $(ee,1)$ case is the solid blue curve.


We also note that this factorization is far from obvious from the context of angles as well.
First, the $\theta_{23}$ and $\delta$ sector can be factored in a straightforward fashion by the Toshev identity \cite{Toshev:1991ku}
\begin{equation}
\sin2\wh\theta_{23}\sin\wh\delta=\sin2\theta_{23}\sin\delta\,.
\end{equation}
This statement is further enhanced by the fact that it is numerically known that $\wh\theta_{23}$ and $\wh\delta$ don't vary much in matter.
After factoring out those two parameters, this leaves $s_{\wh{13}}c_{\wh{13}}^2s_{\wh{12}}c_{\wh{12}}=\wh J/s_{\wh{23}}c_{\wh{23}}\sin\wh\delta$.
This term can be further factored into two terms where one is governed by the atmospheric mass splitting and the other by the solar mass splitting.
For the atmospheric splitting, the factorization is simple
\begin{align}
s_{\wh{13}}c_{\wh{13}}&\approx ~ s_{13}c_{13}\,/\,\Satm\,,\label{eq:atm fact}
\end{align}
Eq.~\ref{eq:atm fact}, follows directly from the zeroth order approximation of DMP \cite{Denton:2016wmg}.

Counterintuitively, for the solar splitting 
\begin{align}
c_{\wh{13}}s_{\wh{12}}c_{\wh{12}}&\approx ~ c_{13}s_{12}c_{12}\, /\, \Ssol\,.\label{eq:sol fact}
\end{align}
The extra $c_{\wh{13}}$ is required on the LHS of eq. \ref{eq:sol fact} to give the LHS the same $a \rightarrow \infty$ limit as the RHS.
There is no direct analog in DMP for eq.~\ref{eq:sol fact}.
Each of these approximations, eqs.~\ref{eq:atm fact}-\ref{eq:sol fact}, are precise at the $0.4\%$ level.
However, when combined, there is a further cancellation and the product has a precision of $0.04\%$.
This returns us to our primary result, eq.~\ref{eq:Jmat}.

\section{Understanding the Precision}
\begin{figure}
\centering
\includegraphics[width=\columnwidth]{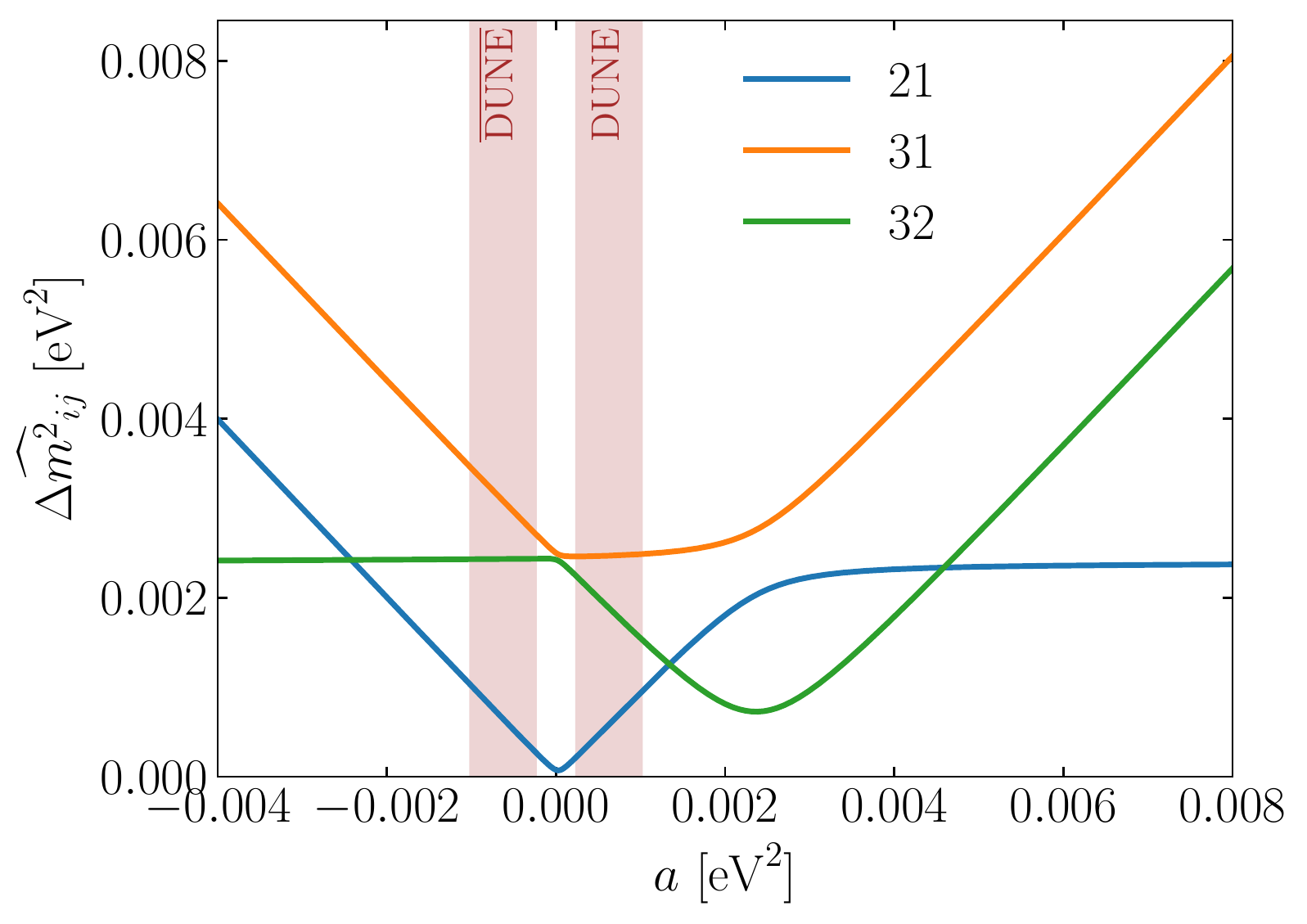}
\caption{The dependence of the three $\Delta\widehat{m^2}$'s in matter for the normal ordering.
DUNE's region of interest for anti-neutrinos and neutrinos are shown in the shaded regions.}
\label{fig:eigenvalues}
\end{figure}

In order to understand why eqs.~\ref{eq:Jmat} and \ref{eq:S13,S12} achieve such precision and to estimate the precision in this factorization analytically, we 
use the exact Naumov-Harrison-Scott (NHS) identity \cite{Naumov:1991ju,Harrison:1999df}, 
\begin{equation}
\wh J\Delta\wh{m^2}_{32}\Delta\wh{m^2}_{31}\Delta\wh{m^2}_{21}=J\Delta m^2_{32}\Delta m^2_{31}\Delta m^2_{21}\,,
\end{equation}
to rewrite the approximate factorization in terms of the exact matter eigenvalues,
\begin{equation}
\Delta\wh{m^2}_{32}\Delta\wh{m^2}_{31}\Delta\wh{m^2}_{21}\approx\Ssol\Satm\Delta m^2_{32}\Delta m^2_{31}\Delta m^2_{21}\,.
\label{eq:approx}
\end{equation}
While the exact eigenvalues have a very complicated analytic form \cite{Zaglauer:1988gz} due to the presence of the $\cos(\frac13\cos^{-1}\cdots)$ terms, we have discovered that the square of the product of the difference of the eigenvalues in matter (also $1/\wh J^2$ thanks to the NHS identity) can be written without any appearance of these $\cos(\frac13\cos^{-1}\cdots)$ factors, see section \ref{sec:exact invariant} below.
In fact, $1/\wh J^2$ can be written as a simple polynomial of the vacuum parameters and the matter potential.
Since each of the $\Delta\wh{m^2}$'s scale with the matter potential in some way as shown in fig.~\ref{fig:eigenvalues}, one would expect that the product of all three, squared, would be sixth order in the matter potential.
In fact, the the product of the three $\Delta\wh{m^2}$'s squared is, in fact, only fourth order.
This statement is independent of the hierarchical measured neutrino mass splittings.
This suggests that only two matter corrections of the form given in eq.~\ref{eq:S13,S12} are needed and clearly justifies the form of eq.~\ref{eq:Jmat}.

Next, to understand the specific terms in $\Satm$ and $\Ssol$, we calculate various limits.
First, consider the large matter potential limit, $a \rightarrow \infty$, then the eigenvalues are as follows:
\begin{align}
\wh{m^2}_3  & \rightarrow    a+\Delta m^2_{ee}s^2_{13}  \nonumber  \\
\wh{m^2}_2  & \rightarrow    \Delta m^2_{31}c^2_{13} \left(1+ {\mathcal O}(\epsilon^2 s^2_{12} c^2_{12}) \right)  \\
\wh{m^2}_1 &   \rightarrow    \Delta m^2_{21}c^2_{12}\left(1+ {\mathcal O} ( \epsilon^2 s^2_{12} c^2_{12}) \right) \,. \nonumber
\end{align}
Thus the $\Delta \wh{m^2}_{jk}$ are 
 \begin{align}
\Delta\wh{m^2}_{31}  & \approx a \approx   \Delta m^2_{21} \Ssol / c^2_{13}  \nonumber   \\ 
\Delta\wh{m^2}_{32}   & \approx a \approx  \Delta m^2_{ee} \Satm    \\
\Delta\wh{m^2}_{21}  &  \approx  c^2_{13}   \left( \frac{\Delta m^2_{31} \Delta m^2_{32} }{\Delta m^2_{ee}}    \right)  \nonumber \\
&  \hspace*{-1.cm}  \times (1+ {\mathcal O}(\epsilon^2 s^2_{12} c^2_{12}) +  {\mathcal O}(\epsilon s^2_{13} \cos2 \theta_{12}) )\,.\nonumber
\end{align}
To understand why we associate $\Delta\wh{m^2}_{31} \approx   \Delta m^2_{21} \Ssol / c^2_{13}$ and  $\Delta\wh{m^2}_{32}   \approx  \Delta m^2_{ee} \Satm$ see \cite{Denton:2018cpu}.  Thus it is simple to see that
\begin{equation}
\Delta\wh{m^2}_{32}\Delta\wh{m^2}_{31}\Delta\wh{m^2}_{21} \approx\Ssol\Satm\Delta m^2_{32}\Delta m^2_{31}\Delta m^2_{21}
\nonumber
\end{equation}
in this limit.  For this the factorization to work, it is crucial that the limit of 
\begin{equation}
\Delta\wh{m^2}_{21}    \approx  c^2_{13}   \left( \frac{\Delta m^2_{31} \Delta m^2_{32} }{\Delta m^2_{ee}} \right) .
\nonumber
\end{equation}
This is where the appearance of the two essential factors, $c^2_{13}$ and $\Delta m^2_{ee}$, come from for this limit. What is highly non-trivial is that it is exactly these factors that are needed at the solar resonance and the atmospheric resonance respectively.

Now consider a number of other simplifying limits:  if $s^2_{13}=0$ and $s^2_{12}=0$ then it is trivial to show the factorization of eq.~\ref{eq:Jmat} is exact.  If $s^2_{13} \neq 0$ and $s^2_{12}=0$ then factional corrections to the factorization are 
$\epsilon s^2_{13}$ whereas  if $s^2_{13} = 0$ and $s^2_{12} \neq 0$ then fractional corrections are $\epsilon^2 s^2_{12} c^2_{12}$, see Supplemental Material.

\section{Exact Jarlskog Invariant in Matter} 
\label{sec:exact invariant} 
For the general cases with   $s^2_{13} \neq 0$ and $s^2_{12} \neq 0$, start from the cubic characteristic equation for the eigenvalues of the square of the neutrino masses in matter, $\wh{m^2}_j$, which each satisfy
\begin{align}
\left(\wh{m^2}_j \right)^3 -A\,\left(\wh{m^2}_j\right)^2+B\,\wh{m^2}_j-C=0\,,
\label{eq:char}
\end{align}
where $A$, $B$, and $C$ are the sum of the eigenvalues, sum of the products of the eigenvalues, and the triple product of the eigenvalues,
\begin{align}
A&\equiv\sum_j \wh{m^2}_j 
=\Delta m^2_{31}+\Delta m^2_{21}+a\,,\nonumber\\
B&\equiv\sum_{j>k}\wh{m^2}_j\wh{m^2}_k
= \label{eq:evaluefn}\\
&\Delta m^2_{31}\Delta m^2_{21}+a(\Delta m^2_{31}c^2_{13}+\Delta m^2_{21}(c^2_{12}+s^2_{13}s^2_{12}))\,,\nonumber\\[2mm]
C&\equiv\prod_j \wh{m^2}_j 
=a \Delta m^2_{31} \Delta m^2_{21} c^2_{13} c^2_{12}\,,\nonumber
\end{align}
using the convention that $(\wh{m^2}_1, \wh{m^2}_2, \wh{m^2}_3 
)=(0,\Delta m^2_{21}, \Delta m^2_{31})$ in vacuum.

Then it is straightforward to show that, 
\begin{multline}
\left(\prod_{j>k}\Delta\wh{m^2}_{jk}\right)^2=\\(A^2-4B)(B^2-4AC)+(2AB-27C)C\,.
\label{eq:Dprod}
\end{multline}
This equation is a general identity for eq.~\ref{eq:char}, independent of the exact values of A, B, C, and is invariant under making the same shift of all of the eigenvalues, as it must given the LHS.

Combining eqs.~\ref{eq:evaluefn} and \ref{eq:Dprod} one obtains the exact expression for  $\left(\prod_{i>j}\Delta\wh{m^2}_{ij}\right)^2$ as a fourth order polynomial in the matter potential, $a$.
This guarantees its factorization in two quadratics as shown by Lodovico de Ferrari in 1540.
To our knowledge this is the only exact measurable expression relating to neutrinos oscillating in matter without the $\cos(\frac13\cos^{-1}\cdots)$ term\footnote{In principle expressions like $A$, $B$, and $C$ in eq.~\ref{eq:evaluefn} are measurable as well, and they don't contain the $\cos(\frac13\cos^{-1}\cdots)$ term. In practice, measuring all three eigenvalues in the \emph{same} matter effect is extremely difficult and not likely to occur even with future experiments. At the moment $\Delta\wh{m^2}_{21}$ is only measured in the sun while $\Delta\wh{m^2}_{31}$ and $\Delta\wh{m^2}_{32}$ are only measured in the Earth making a direct sum or product of these quantities in eq.~\ref{eq:evaluefn} not possible.}.
This means that it is, in principle, possible to write the exact Jarlskog in matter in the same form as eq.~\ref{eq:Jmat}, however since the solutions to the quartic expression which go in to eq.~\ref{eq:S13,S12} are a function of a cubic equation and are extremely complicated, the exact solution is far from simple.
By leveraging the fact that we know that $s_{13}^2$ and $\Delta m^2_{21}/\Dmsqee$ are small numbers, it is possible to drastically simplify that expression to one that is extremely compact.

\section{Error Estimate}
\label{sec:error estimate}
Next, we compare the exact expression and our approximate expression and expand the difference in powers of $\eps$ and $s_{13}^2$.
One can show that the correction $\Delta(\pi^2)$ is well approximated by the simple form,
\begin{align}
\left(\prod_{i>j}\Delta\wh{m^2}_{ij}\right)^2 =&  ~\Ssol^2\,\Satm^2\left(\prod_{i>j}\Delta m^2_{ij}\right)^2 + \Delta (\pi^2)  \label{eq:C11} \\[3mm]
{\rm where} \quad \Delta (\pi^2)   \approx
&+2 \eps^2 s^2_{13} a (\Dmsqee)^5  \label{eq:Dpisq} \\
&-2\eps s_{13}^2\cos2\theta_{12}\,  a^2 \Satm^2 (\Dmsqee)^4 
\nonumber \\
&- 2 \eps^2 s^2_{12}c^2_{12} a^3(\Dmsqee-a) (\Dmsqee)^2 \,.\nonumber
\end{align}
This approximate expression for $\Delta (\pi^2)$ contains the first corrections to the factorization for each power of the matter potential.
There is no constant term ($a^0$) because the approximate expression is exact in vacuum, and there are only terms up to $a^4$ since both the exact and the approximate expressions only have terms up to $a^4$.  An exact expression for $\Delta (\pi^2)$ can easily be obtained using eqs.~\ref{eq:Dprod} and \ref{eq:evaluefn}, see appendix \ref{sec:appA}.
The fractional corrections are of $\mathcal O(\eps s_{13}^2)$ or $\mathcal O(\eps^2 s^2_{12}c^2_{12})$ for each power of $a$ and are of order a few $\times 10^{-4}$.

By propagating the correction from the product of $\Delta\wh{m^2}$'s squared to the correction in $\wh{J}$ via the Naumov-Harrison-Scott identity, we find that the fractional precision in $\wh{J}$ is approximately given by
\begin{align}
\frac{\Delta\wh{J}}{\wh{J}} & \approx         \frac{\Delta (\pi^2)}{2 ~\Ssol^2\,\Satm^2\left(\prod_{i>j}\Delta m^2_{ij}\right)^2 
} 
\label{eq:Jh error}
\end{align}
up to an overall sign.
 In fig.~\ref{fig:error estimate} we plot eq.~\ref{eq:Jh error} (note that using either the exact expression for the denominator or the approximate factorized expression given by eq.~\ref{eq:Jmat}  is indistinguishable).
Also shown for comparison is the exact fractional precision of eqs.~\ref{eq:Jmat} and \ref{eq:S13,S12} as in fig.~\ref{fig:error estimate}.  The agreement between the approximate, eq.~\ref{eq:Jh error}, and the exact fractional correction is excellent.
We note that this precision estimate gets the magnitude of the precision correct as well as the general features: the precision goes to zero for small $a$ and peaks at the level of $0.04\%$.
In addition, the difference passes through zero for $a\approx \Dmsqee$ for neutrinos but not for anti-neutrinos as reasonably expected due to the atmospheric resonance.


To gain further insight, we explore the small and large matter potential limits.
First we evaluate the precision below the solar resonance, $|a|\ll\Delta m^2_{21}\cos 2 \theta_{12}$, and above the atmospheric resonance, $ |a|\gg|\Dmsqee|$.
Using the expression for $\Delta (\pi^2)$, eq.~\ref{eq:Dpisq}, we find that in the low (high) $a$ limit we have that the fractional precision scales like
\begin{align}
\lim_{a\to0}\frac{\Delta\wh J}{\wh J}&\approx \, \eps s_{13}^2  \frac{a}{\Delta m^2_{21} }=3 \e{-4}\frac{a}{\Delta m^2_{21} \cos 2 \theta_{12}}\\
\lim_{a\to\infty}\frac{\Delta\wh J}{\wh J}&\approx \,\eps s_{13}^2 \cos 2 \theta_{12} -\eps^2s_{12}^2c_{12}^2 =6\e{-5}\,,
\label{eq:largea}
\end{align}
up to overall signs.
It is interesting to note that the excellent precision at large $a$ is due to the significant cancellation in eq.~\ref{eq:largea} that happens only for the NO.
For the IO there is no cancellation since $\eps<0$ and the precision levels off at $a\gtrsim\Dmsqee$ to $\Delta\wh J/ \wh J=4\e{-4}$.

\section{Discussion}
\label{sec:discussion}
The Jarlskog invariant in vacuum can be written in a convention independent\footnote{By convention independent, we mean regardless of how one parameterizes the lepton mixing matrix -- that is, without any reference to the mixing angles.} fashion,
\begin{equation}
J=\Im(U_{e1}U_{\mu2}U_{e2}^*U_{\mu1}^*)\,.
\end{equation}
In addition, the Jarlskog in matter must also be convention independent due to the Naumov-Harrison-Scott identity.
Therefore it must be the case that the approximate matter corrections $\mathcal S_{\odot,{\rm atm}}$ can also be written in a convention independent way.
These expressions are,
\begin{align}
\Ssol^2&=\left(1-(1-|U_{e3}|^2)\frac a{\Delta m^2_{21}}\right)^2+4|U_{e2}|^2\frac a{\Delta m^2_{21}}\,,\nonumber\\
\Satm^2&=\left(1-\frac a{\Dmsqee}\right)^2+4|U_{e3}|^2\frac a{\Dmsqee}\,,
\label{eq:indep}
\end{align}
which are essentially as simple as those with mixing angles in eq.~\ref{eq:S13,S12}. See Appendix \ref{sec:appB} for alternative ways to write these resonance factors.
We can similarly write the atmospheric mass splitting as
\begin{equation}
\Dmsqee=\Delta m^2_{31}- \frac{|U_{e2}|^2}{1-|U_{e3}|^2} \Delta m^2_{21}\,, 
\end{equation}
where $|U_{e3}|^2 \approx 0.022$ and $|U_{e2}|^2 \approx 0.31$ and unitarity implies $|U_{e1}|^2=1-|U_{e2}|^2-|U_{e3}|^2$.
It is natural to associate the correction for the atmospheric resonance factor on the $\Delta m^2$ as the minimum separation is $\Delta m^2_{ee} \sin 2\theta_{13}$, whereas for the solar resonance factor, since the minimum separation is not altered at $\Delta m^2_{21} \sin 2 \theta_{12}$, it's natural to associate the correction with the matter potential.

In light of the Naumov-Harrison-Scott identity, it isn't surprising that $\wh{J}/J$ has a form that looks like the inverse of several matter corrections to the $\Delta m^2$'s.
It may not be obvious, however, why $\wh{J}$ is well-approximated by only two such expressions instead of all three.
The reason is because for nearly any value of $a$, there is always one $\Delta\wh{m^2}$ that is essentially constant, see fig.~\ref{fig:eigenvalues}.
In the NO this is $\Delta\wh{m^2}_{32}$ for anti-neutrinos and $\Delta\wh{m^2}_{\ell1}$ for neutrinos where $\ell=3$ below the atmospheric resonance and $\ell=2$ above the atmospheric resonance.
As such having two $\mathcal S$ terms is to be expected.
This fact is further reinforced by the fact that the exact expression for the square of the product of the $\Delta\wh{m^2}_{ij}$ is fourth order in $a$, which directly proves that only two $\Delta\wh{m^2}_{ij}$ require a matter correction.
Moreover, we expect that, when squared, that correction should be fourth order in the matter potential, as is the case for $\Satm^2\Ssol^2$.

In addition, while the presence of the $c_{13}^2$ term breaks an otherwise relatively symmetric definition of $\Ssol$ and $\Satm$, this can be understood using the DMP \cite{Denton:2016wmg} expressions.
In that formalism the (1-2) sector is handled second and thus contains a small (1-3) correction since the (1-3) sector was handled first.


We also note that the functional form of the $\mathcal S$ functions have shown up elsewhere in the literature \cite{Freund:2001pn,Akhmedov:2004ny,Minakata:2015gra,Denton:2018cpu,Wang:2019yfp} with slight differences.
One interesting example is the quantity $\Dmsqee \Satm$ (using the same definition as in eq.~\ref{eq:S13,S12}) which was shown \cite{Denton:2018cpu} to be an excellent approximation for the frequency of $\nu_e$ disappearance in matter, i.e.~$\Delta m^2_{ee}$ in matter.

It is useful to consider the level of precision in a broader phenomenological context.
Since DUNE is striving to reach percent level precision, it is clear that 2\% precision, such as that reached in ref.~\cite{Wang:2019yfp} is not sufficient, thus precision at the 0.04\% level is necessary to be precise enough for DUNE.
On the other hand, there are unavoidable uncertainties in the oscillation probabilities due to uncertainties in the matter density.
A 1\% overall uncertainty in the matter density profile results in a 0.1\% fractional uncertainty in the oscillation probability.
Alternatively, by considering various density profiles available \cite{Roe:2017zdw}, the shape uncertainty can be estimated, and is found to be only 0.01\% \cite{Kelly:2018kmb}.
Thus precision at the 0.04\% level is also a sufficient level of precision.

Finally, we recall the exact Toshev identity, $\sin2\wh\theta_{23}\sin\wh\delta=\sin2\theta_{23}\sin\delta$ \cite{Toshev:1991ku}.
Thus, in terms of the mixing angles, the ratio $\wh J/J$ has no dependence (explicit or otherwise) on $\theta_{23}$ or $\delta$.
This fact is approximately confirmed by the fact that $\Satm$ and $\Ssol$ have no dependence on $\theta_{23}$ and $\delta$.
This brings to attention an important point on CP violation: the quantity that describes the amount of CP violation in the lepton sector is the Jarlskog invariant, $J$, not the CP-violating phase, $\delta$, and thus it is the Jarlskog that should be reported by experiments as a measure of the amount of CP violation.
Since DUNE will measure each part of the vacuum Jarlskog in the atmospheric sector, $\theta_{23}$, $\theta_{13}$, and $\delta$ and could measure $\theta_{12}$ using solar neutrinos as well \cite{Capozzi:2018dat}, it is, in principle, possible to completely determine the Jarlskog in the lepton sector with a single experiment.

\section{Conclusions}
\label{sec:conclusions}
In this paper we have shown that the Jarlskog invariant in the lepton sector is exactly equal to the Jarlskog in matter times two matter resonance factors.
While the coefficients of $a$ and $a^2$ in these resonances are extremely complicated solutions to a quartic equation, they can be simplified immensely while still retaining precision at the $\sim0.04\%$, i.e.~${\mathcal O}(s_{13}^2\cos2\theta_{12} (\Delta m^2_{21}/\Delta m^2_{ee}))$ and ${\mathcal O}(s^2_{12}c^2_{12} (\Delta m^2_{21}/\Delta m^2_{ee})^2)$, a level of precision that is both necessary and sufficient for future long-baseline experiments.

We have also derived the exact Jarlskog invariant in matter as a simple fourth order polynomial in the matter potential which allows us to estimate analytically the precision of the factorization.
To achieve the high precision given here, it is crucial to use the $\theta_{13}$ corrected value for the matter potential for the solar (1-2) sector, $ac_{13}^2$ as well as the effective $\Delta m^2_{ee}$ instead of $\Delta m^2_{31}$ or $\Delta m^2_{32}$ for the atmospheric (1-3) sector.
This precision factorization of the Jarlskog invariant in matter further enhances our understanding of neutrinos in matter relevant for the 
 the currently running NOvA 
 and T2K 
 experiments and the upcoming DUNE 
 and T2HK(K).

\begin{acknowledgments}
PBD acknowledges the United States Department of Energy under Grant Contract desc0012704.
Fermilab is operated by the Fermi Research Alliance under contract no.~DE-AC02-07CH11359 with the U.S.~Department of Energy. 
SP thanks IFT of Madrid for wonderful hospitality that inspired this work.
This project has received funding/support from the European Union’s Horizon 2020
research and innovation programme under the Marie Sklodowska-Curie grant agreement No 690575 and No 674896.
\end{acknowledgments}


\appendix

\section{Exact Derivation}
\label{sec:appA}
\newcommand{\twobma}{( \Delta m^2_{ee} \cos 2 \theta_{13} + \Delta m^2_{21} c^2_{12}) }
\newcommand{\twoacmbb}{( \Delta m^2_{31}  \Delta m^2_{21}) ( \Delta m^2_{31} c^2_{13} \cos 2 \theta_{12} + \Delta m^2_{21} (c^2_{12} c^2_{13}-s^2_{13})) }
\newcommand{\azero}{( \Delta m^2_{31}  + \Delta m^2_{21} ) }
\newcommand{\bone}{( \Delta m^2_{ee} c^2_{13}  + \Delta m^2_{21} ) }
\newcommand{\cone}{ \Delta m^2_{31} \Delta m^2_{21} c^2_{13} c^2_{12} }

First consider the case when $s^2_{13}=0$ with non-zero $s^2_{12}$, then it is easy to show that

\begin{align}
 \left(\prod_{i>j}\Delta\wh{m^2}_{ij}\right)^2 
& =   \biggr(~ (\dmsqss)^2- 2 a \dmsqss \ctot + a^2 ~\biggr) \nonumber \\
& \hspace*{-1.5cm} \times   \biggr(~ \dmsqatm \dmsqatn- a (\dmsqatm -\dmsqss c^2_{12}) ~ \biggr)^2
\end{align}
exactly.
To understand how this result relates to the factorization given by eq.~\ref{eq:Jmat}, one needs the following exact identity 
\begin{equation}
(\dmsqatm - \dmsqss c^2_{12} ) = \frac{ \dmsqatm \dmsqatn+ (\dmsqss)^2 s^2_{12} c^2_{12}} {\dmsqee}\,,
\label{eq:idee}
\end{equation}
where we note that $\dmsqatm - \dmsqss c^2_{12} \approx (\Delta m^2_{\mu \mu}+\Delta m^2_{\tau\tau})/2$ of ref.~\cite{Nunokawa:2005nx}.
If one drops the $(\dmsqss)^2$ terms in this identity then one recovers the $s^2_{13}=0$ limit of the factorization, eq.~\ref{eq:Jmat}.
As we will see, all of the $\epsilon^2 s^2_{12} c^2_{12}$ corrections come from this identity. In principle one could absorb the second term in this identity into the definition of $\dmsqee$ and remove such corrections.

For the case when $s^2_{12}=0$ and non-zero $s^2_{13}$, one can show that without approximation that
\begin{align}
 \left(\prod_{i>j}\Delta\wh{m^2}_{ij}\right)^2 
& =  \biggr(~ (\dmsqee)^2- 2a \dmsqee \cos 2 \theta_{13} + a^2 ~ \biggr)  \nonumber \\
& \hspace*{-2cm} \times  \biggr(~ \dmsqss \dmsqatn - a(c^2_{13}\dmsqatn -s^2_{13} \dmsqss) ~\biggr)^2 .
\end{align}
In this limit $\dmsqee=\dmsqatm$, so that the factorization of eq.  \ref{eq:Jmat} is recovered, if the $a s^2_{13} \dmsqss$ term is set to zero.

Finally, one can show that when $s^2_{13} \neq 0$ and $s^2_{12} \neq 0$, the most general case, that
\vspace{-1cm}
\begin{widetext}
\begin{align}
\left(\prod_{i>j}\Delta\wh{m^2}_{ij}\right)^2 &
  =   \biggr(~ (\dmsqss)^2- 2 a \dmsqss c^2_{13} \ctot + a^2 c^4_{13} ~\biggr) \nonumber \\[1mm]
 &  \times   \biggr(~ (\dmsqatm \dmsqatn)^2- 2 a (\dmsqatm \dmsqatn) (\dmsqatm -\dmsqss c^2_{12})\cos 2 \theta_{13} + a^2 (\dmsqatm -\dmsqss c^2_{12})^2 ~ \biggr)  \nonumber \\[4mm]
  & + ~~(s^2_{13} \dmsqss)~ \biggr(  \sum^4_{n=1} a^n P_n   \biggr)\,. \label{eq:exact}
\end{align}
\end{widetext}
where 
\begin{align}
P_1 &=   2\dmsqss \dmsqatm  \dmsqatn \dmsqee  \label{eq:P1} \\[3mm]
%
P_2 &=   -2\dmsqatm \dmsqatn \dmsqee c^2_{13} \ctot      \label{eq:P2}   \\
& \quad \quad  - 4\dmsqatm \dmsqatn \dmsqss (\cos 2\theta_{13}-c^2_{13} s^2_{12} c^2_{12} )  \nonumber \\
& \quad \quad + (\dmsqee)^2 \dmsqss s^2_{13} 
+2(\dmsqss)^3s^2_{12} c^2_{12} \nonumber \\[3mm]
%
P_3 &=2(\dmsqatm \dmsqatn+(\dmsqee)^2)  c^2_{13} \ctot \cos 2\theta_{13} \nonumber \\
&\quad \quad +4\dmsqee \dmsqss ( s^2_{13} + s^2_{12}  c^2_{12} c^4_{13}  ) \label{eq:P3} \\
& \quad \quad -2(\dmsqss )^2 \cos 2\theta_{12} ( 1+ 2s^2_{12} c^2_{12} c^2_{13})
\nonumber  \\[3mm]
%
%
P_4 &=   - (\dmsqatm+\dmsqatn )\, c^2_{13} \ctot  + \dmsqss .   \label{eq:P4}
\end{align}
Note, that all the corrections to the factorization here are proportional to $s^2_{13} \dmsqss$. $\Delta(\pi^2)$ of eq.~\ref{eq:Dpisq} can now be trivially derived from eqs.~\ref{eq:exact} and \ref{eq:idee}, \ref{eq:P1}-\ref{eq:P4}, as well as the higher order corrections to $\Delta (\pi^2)$.


The zeros of $\left(\prod_{i>j}\Delta\wh{m^2}_{ij}\right)^2$, in the complex matter potential plane, can also be obtained, for both the exact expression and our approximate factorized expression.
Given the exact expressions for the zeros (which are extremely complicated and involve solving a general quartic equation which, in turn, involves solving a general cubic equation) it is possible, in principle, to write an \emph{exact} expression of the form of eqs.~\ref{eq:Jmat} and \ref{eq:S13,S12} with different functions of the vacuum parameters in the $\Satm$ and $\Ssol$ functions.

For the approximate expression, the zeros occur at
\begin{equation}
a= (\Delta m^2_{21}/c^2_{13})\, e^{\pm i 2 \theta_{12}} ~~ {\rm and}  ~~ a=\Delta m^2_{ee}\, e^{\pm i 2 \theta_{13}} \,.
\label{eq:a-zeros}
\end{equation}
\\
Note the appearance of both $c^2_{13}$ and $\Delta m^2_{ee}$ in the location of these zeros. Without these exact factors  the positions of the zeros move by 1 - 2\%.
For the exact expression, we have  numerically calculated the location of the zeros in the complex matter potential plane and find that the fractional corrections  to their position, compared to \ref{eq:a-zeros},  is
\begin{equation}
 6 \times 10^{-4} \quad {\rm and}  \quad 2 \times 10^{-4} \,, \nonumber 
\end{equation}
for the solar and atmospheric zeros, respectively.  For any polynomial, the location of the zeros determines the polynomial exactly, up to an overall factor, therefore the above, again, points to the necessity of including {\it both}  $c^2_{13}$ and $\Delta m^2_{ee}$ in the factorization.

\section{Alternative Expressions for the Resonance Factors}
\label{sec:appB}
Alternatively,  $\Ssol$ can also be written as $\sqrt{1-2\cos2\theta_{12}(c_{13}^2a/\Delta m^2_{21})+(c_{13}^2a/\Delta m^2_{21})^2}$ and $\sqrt{(1-c_{13}^2a/\Delta m^2_{21})^2+4s_{12}^2(c_{13}^2a/\Delta m^2_{21})}$ and similarly for $\Satm$.
Like the Jarlskog invariant, these $\mathcal S$ factors can also be written in a convention independent form, see eq.~\ref{eq:indep}.
They can also be written as $|e^{2i\theta_{12}}-c_{13}^2a/\Delta m^2_{21}|$ and similarly for $\Satm$ which shows where the complex zeros are, see appendix \ref{sec:appA}.

\bibliography{MatterJarlskog}

\end{document}